\documentclass[12pt,a4paper]{article}

\usepackage{amsthm,amsmath,amsfonts,amssymb,natbib}
\usepackage[colorlinks,citecolor=blue,urlcolor=blue]{hyperref}
\usepackage{mathrsfs}
\usepackage{enumerate}

\newtheorem{theorem}{Theorem}
\newtheorem{lemma}{Lemma}
\newtheorem{corollary}{Corollary}
\newtheorem{alg}{Algorithm}
\theoremstyle{definition}
\newtheorem{definition}{Definition}
\newtheorem{example}{Example}

\newcommand{\II}{\mathbb{P}}
\newcommand{\I}{\mathcal{P}}
\newcommand{\PP}{\mathscr{P}}
\newcommand{\MM}{\mathbb{M}}

\newcommand{\HH}{\mathscr{H}}
\newcommand{\CC}{\mathscr{C}}
\newcommand{\M}{\mathscr{M}}
\newcommand{\FF}{\mathscr{F}}

\newcommand{\EE}{\mathsf{E}}
\newcommand{\XX}{\mathsf{X}}

\newcommand{\order}[1]{\langle\langle #1\rangle \rangle}

\begin{document}
\title{The Gibbs Sampler Revisited from the Perspective of Conditional Modeling}
\author{Kun-Lin Kuo\\
    \small Institute of Statistics, National University of Kaohsiung, Kaohsiung, Taiwan\\
     and \\
    Yuchung J. Wang\thanks{Corresponding author: yuwang@camden.rutgers.edu}\\
    \small Department of Mathematical Sciences, Rutgers University, Camden, NJ, USA}
\date{}
\maketitle

\begin{abstract}
The Gibbs sampler (GS) is a crucial algorithm for approximating complex calculations, and it is justified by Markov chain theory, the alternating projection theorem, and $I$-projection, separately. We explore the equivalence between these three operators.
Partially collapsed Gibbs sampler (PCGS) and pseudo-Gibbs sampler (PGS) are two generalizations of GS.
For PCGS, the associated Markov chain is heterogeneous with varying state spaces, and we propose the iterative conditional replacement algorithm (ICR) to prove its convergence. In addition, ICR can approximate the multiple stationary distributions modeled by a PGS.
Our approach highlights the benefit of using one operator for one conditional distribution, rather than lumping all the conditionals into one operator.   Because no Markov chain theory is required, this approach simplifies the understanding of convergence.
\end{abstract}

\noindent%
\vfill

\section{Introduction}
The Gibbs sampler (GS) is one of the most useful algorithms for using computing power to approximate analytically difficult calculations; traditionally, its justification  relies on the Markov chain theory.   \citet{Diaconis2010} showed that iterated conditional expectation \citep{Burkholder1961} is intimately connected to the Gibbs sampler.  In fact, the successive steps in the GS are equivalent to iterated projections in the Hilbert space; this connection allows the alternating projection theorem of \citet{Neumann1950} to be brought in. \citet{Kuo2019} used the minimization of Kullback-Leibler information divergence to study the GS, and named it $I$-projection. The advantages of such an approach include (i) the outcome of $I$-projection has a simple closed form; (ii) it requires little background knowledge to comprehend, and (iii) convergence in Kullback-Leibler divergence implies convergence in total variation norm.

Major generalizations of the GS occur in two directions: the partially collapsed Gibbs sampler \citep[PCGS,][]{vanDyk2008} and the pseudo-Gibbs sampler \citep[PGS,][]{Heckerman2000}. For $\XX=(X_1,\ldots,X_d)$, the conditional distribution involving all $d$ components of $\XX$ is a full conditional; otherwise, it is a non-full conditional.   When all the conditional distributions are derived from the same joint distribution, these conditionals are said to be compatible; otherwise, they are incompatible. The original GS uses exclusively compatible and full conditionals,  while PCGS uses a compatible mixture of full and non-full conditionals to sample. On the other hand, PGS employs an incompatible mixture of full and non-full conditionals;
it is also known as potentially incompatible Gibbs sampler (PIGS), and is widely used in missing data imputation with chained equation.

In this paper, we show:
\begin{enumerate}[(i)]
  \item The Markov operator, the conditional expectation operator, and the $I$-projection are the same operators for full conditionals.
  \item The Markov chain of a PCGS is heterogenous with varying state spaces.   We propose the iterative conditional replacement algorithm (ICR) to prove its convergence.
  \item We will show how ICR computes the  multiple stationary distributions modeled by a PGS.
\end{enumerate}
In addition, a concept called mutually stationary distributions is discussed; the phenomenon happens naturally for PGS.  Whether the joint distribution modeled by a collection of conditional distributions is unique or not, ICR will compute them.

\section{Three ways to justify the Gibbs samplers}
We first consider the two-component GS.  Let random variables $(X,Y)$ have a joint pdf (probability density function) $f(x,y)$ and marginal pdf $f_1(x)$ and $f_2(y)$, with supports $S$, $S_1$, and $S_2$, respectively.  Following the calculus of probabilities, the conditional pdf of $X$ given $Y$ is $f_{1|2}(x|y)=f(x,y)/f_2(y)$ and $f_{2|1}(y|x)$ is similarly defined.  Determining  the joint pdf from a collection of conditional pdfs is called conditional modeling.
When both $f_{1|2}$ and $f_{2|1}$ are easy to sample from but $f$ is not, GS generates samples of the joint using the following kernel:
\[ k(x,y; x^{\ast},y^{\ast})= f_{2|1}(y^{\ast}|x)f_{1|2}(x^{\ast}|y^{\ast}) \]
There are three pathways to  the convergence of the GS, though on the surface they look unrelated.

\subsection{Justify GS using Markov chain}
Let $\HH$ be the space of joint pdfs over $S$.
$\HH_1$ and $\HH_2$ be the spaces of  $X$-marginal pdfs over $S_1$ and $Y$-marginal pdfs over $S_2$, respectively.
Let $K$ be the Markov operator on $\HH$ associated with kernel $k$.  If $f(x,y)$ is absolutely continuous with respect to the Lebesgue measure, $S=S_1 \times S_2$, and the sample chain is aperiodic, then (see Theorem 7.1.14 of \citet{Robert1999}) as $n\to\infty$,
\begin{equation}\label{mk1}
\|K^{n}(x,y;\bullet,\bullet) -f (\bullet,\bullet) \|_{tv} \to 0, \mbox{ for } (x,y) \in S,
\end{equation}
where $\|\bullet\|_{tv}$ is the total variation norm.  This justification is the most popular and is also the reason that GS belongs to the toolbox of MCMC (Markov chain Monte Carlo).  For a $h \in \HH$, it would be helpful to know $K(h)$ in closed form.

\subsection{Justify GS using alternating projections}
Define  $\CC_1=\{ h \in \HH: h_{1|2}=f_{1|2} \}$  and $\CC_2=\{ h \in \HH: h_{2|1}=f_{2|1} \}$.
When $S = S_1 \times S_2$, the unique joint pdf determined by $f_{1|2}$ and $f_{2|1}$ exists in $\CC_1 \cap \CC_2$, which is hard to locate. \citet{Neumann1950} proved the following theorem \citep{Diaconis2010} to approximate the joint pdf:

\begin{theorem}
Let $\PP_1$ and $\PP_2$, respectively, be orthogonal projections onto closed subspaces
$\M_1$ and $\M_2$ of a Hilbert space $\HH$. Let $\PP_{I}$ be the orthogonal projection onto the intersection $\M_1 \cap \M_2$.
If $\PP_{21}=\PP_2 \PP_1$, then $\PP_{21}^k \to \PP_{I}$ as $k \to \infty$. That is, $\|\PP^k_{21}(h) -\PP_{I}(h) \|_{tv}\to 0$ for each $h \in \HH$.
\end{theorem}

The rational is that both $\PP_1$ and $\PP_2$ are easy to execute, then the difficult $\PP_{I}$ can be approximated by repeatedly doing $\PP_{21}$.   Moreover, alternating projection of $h \in \HH$ onto $\CC_1$ is realized by the conditional expectation operator, that is, $\PP_1(h)=\EE[h \mid\CC_1]= \EE[h\mid f_{1|2}]$. \citet{Diaconis2010} demonstrate (i) $\PP_1$ is a contraction operator in $\HH$ in terms of the total variation norm, $\|\bullet\|_{tv}$; and (ii) the Markov operator $K(\bullet)$ is equivalent to $\EE [\EE[\bullet \mid\CC_2]\mid\CC_1]:= \PP_{21}$.  Formally, for any $h \in \mathscr{H}$ and $(x,y) \in S$,  we can show that
\begin{equation}\label{mk2}
K(h)(x,y)= \PP_{21}(h) (x,y).
\end{equation}
Now, we cannot help wanting to know the explicit form of $\EE[h\mid \CC_2]$, because it can help numerical implementation of $\PP_{21}$.

\subsection{Justify GS based on $I$-projection}
For any $q \in \HH$ and $h \in \CC_1$, the following equality holds:
\begin{eqnarray*}
I(q;h)&=&\int q \log \frac{q}{h}\\
&=&\int q \log \frac{q}{f_{1|2} q_2} +\int q \log \frac{f_{1|2} q_2}{f_{1|2} h_2}\\
&=& I(q; f_{1|2} q_2) + I (q_2; h_2),
\end{eqnarray*}
where $I(q;h)$ is the Kullback-Leibler divergence between $q$ and $h$.
Hence, $\min_{h \in \CC_1} I(q; h)= I(q; f_{1|2} q_2)$, where $q_2$ and $h_2$ is the $Y$-marginal pdf of $q$ and $ h$, respectively.  We call $f_{1|2} q_2$ the $I$-projection of $q$ onto $\CC_1$.  Denote $I_1$ and $I_2$ as the $I$-projections onto $\CC_1$ and $\CC_2$, respectively.  The closed form shows the effect of $I$-projection; it replaces the $(X|Y)$-conditional pdf of $q$ by the conditional pdf that defines $\CC_1$, while keeps the $Y$-marginal pdf of $q$ unchanged.   

Following the rational of \citet{Neumann1950}, will cyclic $I$-projections between $\CC_1$ and $\CC_2$ lead to projection onto their intersection? \citet{Kuo2019} proved this is true: for any $q^{(0)} \in \HH$,
\[ \lim_{n \to \infty} [I_1 I_2]^n (q^{(0)})= f \in \CC_1 \cap \CC_2. \]
The above alternating projections can be detailed as follows: for $n=0,1,\ldots$,
\[q^{(2n+1)}=I_2 (q^{(2n)})=f_{2|1}q_1^{(2n)}\]
and
\[q^{(2n+2)}=I_1 (q^{(2n+1)})=f_{1|2}q_2^{(2n+1)},\]
with both $q^{(2n)}\to f$ and $q^{(2n+1)}\to f$ as $n\to\infty$.
Moreover, when $\CC_1 \cap \CC_2 =\emptyset$, or equivalently $f_{1|2}$ and $f_{2|1}$ are incompatible, \citet{Kuo2019} showed that $q^{(2n+1)}\to  \pi^{(1,2)} \in \CC_2$ and $q^{(2n)}\to \pi^{(2,1)} \in \CC_1$.
In other words, $(I_1I_2)^n(q^{(0)})$ and $(I_2I_1)^n(q^{(0)})$ converge to different stationary distributions under suitable assumptions.
Notice, the sequential order of $I$-projections is reflected in the superscript of $\pi$.  The following lemma shows how to use the stationary distributions to verify the compatibility of a conditional model.

\begin{lemma}\label{lem:compatible}
Conditional pdfs, $\{ f_{1|2}, f_{2|1} \}$, are derived from the same joint pdf, called them compatible, if $\pi^{(1,2)}= \pi^{(2,1)}$.
\end{lemma}
\begin{proof}
$\pi^{(1,2)} =\pi^{(2,1)}$ implies ${\CC}_1 \cap {\CC}_2 \ne \emptyset$, thus compatible.
\end{proof}
The inverse of Lemma~\ref{lem:compatible} is also true, but the proof requires  additional facts, i.e., $\pi^{(1,2)}_i=\pi^{(2,1)}_i$ for $i=1,2$, see \citet{Kuo2019}.

\subsection{The equivalence of Markov operator, the conditional expectation operator and $I$-projection}\label{sec:2-4}
For $h \in \HH$, the equivalence between $K(h)$ in Eq.~(\ref{mk1}) and $\EE[\EE[h\mid\CC_2]\mid\CC_1]$ has been established in Eq.~(\ref{mk2}); we need to prove that $\EE [h\mid\CC_1]$ and $I_1(h)$ are the same thing.  The definition \citep[p.~313]{Chung2001} of conditional expectation of $Y$ relative to Borel sub $\sigma$-algebra $\xi$ is
\begin{enumerate}[(i)]
\item $\EE[Y\mid \xi]$ is $\xi$-measurable; and
\item $\EE[Y\mid \xi]$ has the same integral as $Y$ over any set of $\xi$.
\end{enumerate}
Let $\FF_1$ and $\FF_2$ be the $\sigma$-algebra on the support of $X$ and $Y$, respectively.  The mapping from $\CC_1$ to $\HH_2$ is bijective, because every $f_{1|2} q_2$ is mapped to a unique $q_2$.   For $h \in \HH$, $I_1(h)=f_{1|2} h_2$ is $(-\infty, \infty) \times \FF_2$-measurable. And for any $a_2, b_2 \in \mathbb{R}$, we have
\begin{eqnarray*}
\int_{a_2}^{b_2} \int_{-\infty}^{\infty} h(x,y) dx dy &=& \int_{a_2}^{b_2} h_2(y) dy\\
&=& \int_{a_2}^{b_2} \int_{-\infty}^{\infty} I_1(h)(x,y) dx dy.
\end{eqnarray*}
Thus, $I_1(h)$ is the conditional expectation of $h$ relative to $(-\infty, \infty) \times \FF_2$.  The equivalence between $I$-projection and the Markov operator $K$ can be directly demonstrated for discrete $X$ and $Y$.   Let $T_1$ be  the transition matrix representing  $f_{1|2}$.   Then, for $q \in \HH$, $qT_1$ is equal to $f_{1|2}(x|y) q_2(y)$, which is $I_1(q)$.  Because
the $K$ in Eq.~(\ref{mk2}) is a combination of $I_1$ and $I_2$, the conditional replacement becomes less obvious.

\begin{lemma}
There is only one pathway to the convergence of the Gibbs sampler because of
\[ K(h)=\EE[\EE[h\mid\FF_2]\mid\FF_1]=I_1 I_2(h) \]
for any $h \in \HH$.
\end{lemma}

When the marginal pdf can be computed such as discrete random variables, the equivalence can be used to implement the iterates of conditional expectation numerically.
Table~\ref{tab1} clarifies different pdf spaces and associated $\sigma$-algebra.
In the following, we will use the equivalence to investigate two variations of GS: either the given conditional pdfs are incompatible ($\bigcap_j \CC_j=\emptyset$) or some of the conditional pdfs are not full conditionals.   For $(x_1,x_2,x_3)$, $\{ f_{1|2}, f_{3|1}, f_{2|13}\}$ has two non-full conditionals and one full conditional pdf, $f_{2|13}$.
\begin{table*}
\caption{Function spaces of the joint, the marginal and the conditional pdfs and associated $\sigma$-algebras}\label{tab1}
\begin{tabular}{ccccc}
\hline
\multicolumn{1}{c}{Name of distribution} & \multicolumn{1}{c}{pdf} & \multicolumn{1}{c}{support} & \multicolumn{1}{c}{function space} & \multicolumn{1}{c}{$\sigma$-algebra}\\
\hline
$X$-marginal & $f_1(x)$ & $S_1$ & $\HH_1$ & $\FF_1$\\
$Y$-marginal & $f_2(y)$ & $S_2$ & $\HH_2$ & $\FF_2$\\
joint & $f(x,y)$ & $S$ & $\HH$ & $\FF_1\times \FF_2$\\
$(X|Y)$-conditional & $f_{1|2}(x|y)$ & $S$ & $\CC_1\subset \HH$ & $(-\infty,\infty)\times \FF_2$\\
$(Y|X)$-conditional & $f_{2|1}(y|x)$ & $S$ & $\CC_2\subset \HH$ & $\FF_1\times (-\infty,\infty)$\\
\hline
\end{tabular}
\end{table*}

\section{Partially collapsed Gibbs sampler}
\subsection{Introduction of PCGS}
Now consider the $d$-dimensional case. Let the $d$ random variables be represented by $\XX=(X_1,\ldots,X_d)$ with joint pdf $f=f(x_1,\ldots,x_d)$.  We shall use $D=\{1,\ldots,d\}$ to denote $\XX$ and $a_i\subset D$ to denote $(X_j: j \in a_i)$, $f_{a_i}$ to indicate the $a_i$-marginal pdf and $f_{a_i|b_i}$ as the conditional pdf of $(X_j: j \in a_i)$ given $(X_k: k \in b_i)$.   The goal is to derive the joint pdf, $f$, from a collection of $L$ conditional pdfs: $\{ f_{a_i|b_i}: 1 \le i \le L \}$.   From $f$ to $\{ f_{a_i|b_i} \}$ is called ``calculus of probability,'' while the GS attempts to solve the inverse problem: using $\{ f_{i|-i}: i \le i \le d \}$, or more generally $\{ f_{a_i|b_i}: 1 \le i \le L \}$, to recover $f$, where $-i=D \backslash \{i\}$.
When $a_i$ is not a singleton, it is called blocking. \citet{Liu1994} argued that ``grouping (some researchers call it blocking) highly correlated components together in the Gibbs sampler can greatly improve its efficiency.''  Blocking fastens the exploration of the support, but turns a univariate sampler (using $f_{i|-i}$) into a multivariate sampler (using $f_{a_i|-a_i}$).  To remedy the side effect of blocking, \citet{vanDyk2008} invented PCGS and used $\{ f_{i|b_i}: 1 \le i \le d\}$ to generate samples of $f$.  The name is due to  collapsing some blocked full conditional pdfs, $f_{a_i|b_i}$, with $a_i\cup b_i=D$, into $f_{i|b_i}$, a non-full conditional pdf.
They also noticed that the updating order (the sequence that components of $\XX$ are generated) has restrictions, unlike the GS where  any of the $d!$  updating orders is valid.

\subsection{The conditional replacement operator for PCGS}
In Section~\ref{sec:2-4}, conditional expectation is equivalent to replacing its conditional pdf and keep its marginal pdf.  That is , $\EE(q|f_{a_j|b_j})=f_{a_j|b_j} q_{b_j}$ if and only if $q_{b_j}$ can be obtained from $q$.
For $\{f_{a_i|b_i}:1\leq i\leq L\}$, let $c_i=a_i \cup b_i$, for every $i$.  Let $\HH$ and $\HH_{c_i}$ be the space of joint pdf and $c_i$-marginal pdf with supports $S$ and $S_{c_i}$, respectively. 
When $c_j \not \subseteq c_i$, Kullback-Leibler divergence between a $q \in \HH_{c_i}$ and a $h \in \HH_{c_j}$ cannot be defined but conditional expectation of $q$ with respect to $\CC_{c_i}$ (more precisely, the $\sigma$-algebra on $S_{c_i}$) may be defined, where
$\CC_{c_i}=\{ h \in \HH_{c_i}: h_{a_i|b_i}=f_{a_i|b_i} \}$.
The  following marginalization operator $\MM$ is needed for such an operation.  Whenever $u \subset v \subseteq D$, $h_u \in \HH_u$ can be derived from $h_v \in \HH_v$ by integration, and
operator $\MM_u$ is define as follows:
\[
\MM_u(h_v)=\int h_v ( \prod_{j\in v\backslash u} dx_j) = h_u.
\]
Then the conditional expectation of $h \in \CC_{c_i}$ with respect to $\CC_{c_j}$ is
\[
\II_i^{j}(h)= f_{a_j|b_j} \MM_{b_j} (h)=f_{a_j|b_j} h_{b_j},
\]
 if and only if, $b_j \subset a_i \cup b_i =c_i$.   Because $\II_i^{j} (h)$ replaces the $h_{a_i|b_i}$ of $h$ with the $f_{a_j|b_j}$ that defines $\CC_{c_j}$, we shall name it the {\it conditional replacement operator}, and it is the basic component that constitutes the algorithm.  In the following, we define the sequences under which $\II_i^{j}$ can operate  in a cyclical fashion.

\subsection{Permissible updating cycles for the conditional replacement operators}
\citet{vanDyk2008} invented a four-step procedure to determine the valid updating cycles. \citet{Kuo2019} proposed a nested condition for PCGS.  Following is the definition of a permissible updating cycle.

\begin{definition}
Given ${\cal A}=\{ f_{a_i|b_i}: 1 \le i \le L\}$ and $c_i=a_i \cup b_i$.
Let $(1^{\ast},\ldots, L^{\ast})$ be a permutation of $(1,\ldots,L)$ with $(L+1)^{\ast} \equiv 1^{\ast}$.
When $b_{(i+1)^{\ast}} \subset c_{i^{\ast}}$ for every $i$, then
$(1^{\ast},\ldots, L^{\ast})$  is said to be a permissible updating cycle for ${\cal A}$, and it is abbreviated as $\order{1^{\ast},\ldots, L^{\ast}}$.
\end{definition}

For $\{ f_{i|-i}: 1 \le i \le d \}$, every permutation of $(1,\ldots,d)$ is  permissible because $c_i=D$ for every $i$, while $\{f_{1|2}, f_{3|1},\\ f_{2|13} \} $ has only one: $\order{1,3,2}$.  Also, $\order{1,3,2}$ is  the only permissible updating cycle for  $\{f_{1|2}, f_{2|3}, f_{3|1} \} $;
this conditional model has no full conditional but cyclic conditional replacements are permitted and the iterations will produce three two-dimensional margins pdfs, see Example~\ref{ex:2} below.
Condition $\order{1^{\ast},\ldots, L^{\ast}}$ enables conditional replacements to complete the cycle:  starting from $\HH_{c_{1^{\ast}}}$, traveling through $\HH_{c_j}, j>1$ until $\HH_{c_{L^{\ast}}}$, then coming back to $\HH_{c_{1^{\ast}}}$. Equivalently, $\EE[\EE[\bullet \mid \CC_{i^{\ast}}]\mid \CC_{(i+1)^{\ast}}]$ is valid because when $b_{(i+1)^{\ast}} \subset c_{i^{\ast}} $ the $\sigma$-algebra over the support of $f_{c_i^{\ast}}$ can be reduced into the correct $\sigma$-algebra over the support of  $f_{b_{(i+1)^{\ast}}}$.

The convergence of the following algorithm will be proved in a broader context in the next section.  The basic idea is that every $\II_{i^{\ast}}^{(i+1)^{\ast}}$ is a contraction operator, then their composite,  $\II_{(L-1)^{\ast}}^{L^{\ast}}\cdots \II_{1^{\ast}}^{2^{\ast}}(h)$, is also a contraction operator.

\begin{alg}[Iterative conditional replacement for PCGS]\label{alg:1}
Let $f$ be the joint pdf of $\XX$.   Given ${\cal A}= \{ f_{a_i|b_i}: 1 \le i \le L\}$ and
assume $\order{1,\ldots,L}$
is a permissible updating cycle for ${\cal A}$.  Assume $a_{L} \cup b_{L}=D$.  Iterative conditional replacement in the order of $a_{1} \to \cdots\to a_{L}$ will converge to $f$ from any initial pdf
$q^{(0)}\in \HH_{c_{L}}$.  Define
$\I=\II_{L-1}^L \cdots \II_1^2$,  as $n \to \infty$,
\[
\I^{n}(q^{(0)}) \to f\mbox{ in Kullback-Leibler divergence.}
\]
\end{alg}

\citet{vanDyk2008} used $\{ f_{i|b_i}: 1 \le i \le d\}$ to generate one component  of $\XX$ at a time.  Their justifications are based on the following line of reasoning:
\begin{enumerate}[(i)]
  \item Turning $\{ f_{i|-i}: 1 \le  i \le d\}$ into  blocked $\{ f_{a_i|b_i}: a_i\cup b_i=D, 1 \le i \le d\}$ does not change the stationary pdf $f$;
  \item When $i < j$ and $a_i \cap a_j \ne \emptyset$, $f_{a_i|b_i}$ can be reduced to $f_{(a_i \backslash a_j )|b_i}$ without changing the stationary distribution;
  \item Iterating (ii) a few times and the original $\{ f_{i|-i}: 1 \le  i \le d\}$  becomes  $\{f_{i|b_i}: 1 \le  i \le d\}$.
\end{enumerate}
These are correct heuristic arguments because rigours proof based on Markov chain theory can be quite complicated.  One reason is  that the state space, $\{i\}\cup b_i$, varies with $f_{i|b_i}$, thus, change the homogenous Markov chain of GS into a heterogeneous chain of PCGS.  

\section{Iterative conditional replacement for pseudo-Gibbs sampler (PGS)}
\subsection{Introduction to dependence network and PGS}
Conditionally specified model $\{f_{a_i|b_i}: 1 \le i \le L\}$ can be regarded as a graphical model composed by $L$ directed graphs; within each graph, a directed edge is initiated from every $j \in b_i$ pointing to every $k \in a_i$.  Set $b_i$ is call the ``parent'' set of $a_i$, so  $f_{a_i|b_i}$ can be written as $f( a_i | PR_{a_i})$.  Estimating each $f_{a_i|PR_{a_i}}$ from data, \citet{Heckerman2000} called the collection: $\{f_{a_i|PR_{a_i}}\}$  a {\it dependence network}, and posted it as an unsupervised learning problem.  Similar conditional modeling for the joint distribution of missing data also appeared in multiple imputation, see \citet{Buuren2018}.
Because $f_{a_i|PR_{a_i}}$ is not derived from a joint pdf but estimated from data,  $\{f_{a_i|PR_{a_i}}\}$ will not be compatible. However, \citet{Heckerman2000} still used GS to approximate the joint pdf.
They coined the term pseudo-Gibbs sampler for the GS based on incompatible conditional pdfs, which is also known as PIGS in multiple imputation.
The first step of PGS is to expand every non-full conditional pdf, say $f_{a_i|PR_{a_i}}$, into a full conditional pdf $f_{a_i| -a_i}$.  Such  expansions can confuse the determination of permissible updating cycles.
\citet{Heckerman2000} argued that when the data are large, $\{f_{a_i|PR{a_i}}\}$ would be nearly compatible, and the approximated joint pdfs with different updating cycles would be nearly identical.
Such a claim has been refuted by many statisticians.
For example, \citet[p.~257]{Casella1996} stated ``Gibbs samplers based on a set of densities that are not compatible result in Markov chains that are \emph{null}, that is, they are either null recurrent or transient.''
\citet[p.~268]{Gelman2001} also stated ``the simulations  never converge to a single distribution, rather the distribution depends upon {\it the order of the updating} and when the updating stopped'' (at which ${\CC}_i$).
That is why \citet[p.~267]{Besag2001} stated that PGS's
``theoretical properties are largely unknown and no doubt considerable caution must be exercised.''
Our thesis will vindicate the claim by \citet{Gelman2001} and completes the theoretical developments mentioned by \citet{Besag2001}.  The following algorithm computes the stationary pdfs for  incompatible $\{f_{a_i|b_i}\}$, but also for compatible $\{ f_{a_i|b_i} \}$.  Thus, Algorithm~\ref{alg:1} is a special case of Algorithm~\ref{alg:2}:

\begin{alg}[Iterative conditional replacement for PGS]\label{alg:2}
Let $\order{1,\ldots,L}$ be a permissible updating cycle for $\{f_{a_i|b_i}: 1\le i \le L\}$ and $c_i=a_i \cup b_i$.
Initiated from any $q^{(0)}\in \HH_{c_{L}}$, define the terms of $L$ sequences: \[\{q^{(Lk+1)}:k=0,1,\ldots\},\ldots,\{q^{(Lk+L)}:k=0,1,\ldots\}\]
by
\[
q^{(Lk+1)}=\II_{L}^1(q^{(Lk)}),\
q^{(Lk+2)}=\II_{1}^2(q^{(Lk+1)}),
\ldots,\]
\[q^{(Lk+L)}=\II_{L-1}^L(q^{(Lk+L-1)}).
\]
Then, each sequence, say $\{q^{(Lk+i)}:k=0,1,\ldots\}$, converges to a stationary pdf in ${\CC}_{c_i}$.

When every $f_{a_i|b_i}$ is derived from the same joint pdf $f$, $\{q^{(Lk+j)}\}$ converges to $f$ whenever $f_{a_{j}|b_{j}}$ is a full conditional.
For non-full  $f_{a_{j}|b_{j}}$, $\{q^{(Lk+j)}\}$ converges to $f_{c_{j}}$.  The convergence is in terms of Kullback-Leibler divergence.
\end{alg}

We first show that operator $\II_i^{i+1}$   is a contraction operator whenever the operation is permissible.
\begin{lemma}\label{lem:shorten}
Assume that $\order{1,\ldots,L}$ is a permissible updating cycle for $\{f_{a_i|b_i}: 1\le i \le L\}$ and $c_i=a_i \cup b_i$.
Mapping any two pdfs $h$ and $g$ in $\CC_{c_i}$,
by $\II_i^{i+1}$ onto $\CC_{c_{i+1}}$ decreases their mutual Kullback-Leibler divergence.
That is,  $I(h;g)>I(\II_i^{i+1} (h); \II_i^{i+1}(g))$.  This contraction holds for every neighboring pair $(i, i+1)$ of $\order{1,\ldots,L}$ with $L+1\equiv 1$.
\end{lemma}
\begin{proof}
Assume $u_i=c_i\backslash b_{i+1} \ne \emptyset$ and $x_{c_i}$ denotes $(x_k, k \in c_i)$ with $x_{u_i}, x_{b_{i+1}}$  similarly defined.  We have
\begin{eqnarray*}
\lefteqn{I(h;g)}\\
&=&\int h\log\frac {h}{g}\,dx_{c_i}\\
&=&\int h_{b_{i+1}}\left[\int h_{u_i|b_{i+1}}\left(\log\frac {h_{u_i|b_{i+1}}}{g_{u_i|b_{i+1}}}+\log\frac{h_{b_{i+1}}}{g_{b_{i+1}}}\right)\,dx_{u_i}\right]\,dx_{b_{i+1}}\\
&=&\int h_{b_{i+1}}I(h_{u_i|b_{i+1}};g_{u_i|b_{i+1}})\,dx_{b_{i+1}} +I(h_{b_{i+1}};g_{b_{i+1}}).
\end{eqnarray*}
It is easy to see that
$I(\II_i^{i+1}(h);\II_i^{i+1}(g))=I(h_{b_{i+1}};g_{b_{i+1}})$, because
$\II_i^{i+1}(h)=f_{a_{i+1}|b_{i+1}} h_{b_{i+1}}$ and $\II_i^{i+1}(g)=f_{a_{i+1}|b_{i+1}} g_{b_{i+1}}$.
Hence,
\begin{eqnarray*}
\lefteqn{I(h;g)-I(\II_i^{i+1}(h);\II_i^{i+1}(g))}\\
&=&\int h_{b_{i+1}}I(h_{u_i|b_{i+1}};g_{u_i|b_{i+1}})\,dx_{b_{i+1}},
\end{eqnarray*}
which is strictly positive, unless $h_{u_i|b_{i+1}} = g_{u_i|b_{i+1}}$.
When $c_i=b_{i+1}$ for all $i$, then $c_L=(\bigcup_{j=1}^{L} a_j) \cup b_1$ violates  $L+1\equiv 1$, and $\order{1,\ldots,L}$ will not be a permissible updating cycle.  Thus, $u_i \ne \emptyset$  for some $i$.
\end{proof}

Contraction operator is usually defined within one metric space or between two metric spaces; here $\II_i^{i+1}$ operates between spaces of marginal distributions of different subsets of $\XX$, and Kullback-Leibler divergence replaces metric.  The following theorem proves that the $L$ sequences
of distributions produced by Algorithm~\ref{alg:2} converge, respectively, to the $L$ stationary distributions in each $\CC_{c_i}$.   When dealing with incompatible models, order of updating plays a critical role, thus every stationary distribution is indexed by it.

\begin{theorem}\label{thm:convergence}
Assume that $\order{1,\ldots,L}$ is a permissible updating cycle for $\{f_{a_i|b_i}: 1\le i \le L\}$ and $c_i=a_i \cup b_i$.
Also assume that the $L$ stationary pdfs: $\pi^{(a_{i+1},\ldots,a_L,a_1,\ldots,a_i)}$, $1 \le i \le L$,
exist.\footnote{This assumption is necessary, see Example~\ref{ex:1} below, because $\CC_{c_i}$ is not a Banach space yet.}  For every $1 \le i \le L$,
the sequence of distributions produced by Algorithm~\ref{alg:2}, $\{q^{(Lk+i)}:k=0,1,\ldots\}$,  converges monotonically to $\pi^{(a_{i+1},\ldots,a_L,a_1,\ldots,a_i)}$ in Kullback-Leibler divergence,
as $k$ tends to $\infty$.
\end{theorem}
\begin{proof}
Due to Lemma~\ref{lem:shorten}, we have, for $1 \le i \le L$,
\begin{eqnarray*}
\lefteqn{I\left(\pi^{(a_{i+1},\ldots,a_L,a_1,\ldots,a_{i})};q^{(Lk+i)}\right)}\\
  &>& I\left(\II_i^{i+1}(\pi^{(a_{i+1},\ldots,a_L,a_1,\ldots,a_{i})});\II_i^{i+1}(q^{(Lk+i)})\right)\\
 &= & I\left(\pi^{(a_{i+2},\ldots,a_L,a_1,\ldots,a_{i+1})};q^{(Lk+i+1)}\right).
\end{eqnarray*}
Applying $\I\equiv \II_{i-1}^i  \cdots \II_{i+1}^{i+2} \II_{i}^{i+1}$ to $q^{(Lk+i)}$, the iterations will travel through $\CC_{a_j}$, $j \ne i$ and return to ${\CC}_{c_i}$ with
$\I(\pi^{(a_{i+1},\ldots,a_L,a_1,\ldots,a_{i})})
= \pi^{(a_{i+1},\ldots,a_L,a_1,\ldots,a_{i})}$,
and $\I (q^{(Lk+i)})= q^{(L(k+1)+i)}$. Thus,
\begin{eqnarray*}
\lefteqn{I\left(\pi^{(a_{i+1},\ldots,a_L,a_1,\ldots,a_i)};q^{(Lk+i)}\right)}\\
&>&
I\left(\I(\pi^{(a_{i+1},\ldots,a_L,a_1,\ldots,a_{i})});\I (q^{(Lk+i)})\right)\\
&=&
I\left(\pi^{(a_{i+1},\ldots,a_L,a_1,\ldots,a_{i})};q^{(L(k+1)+i)} \right).
\end{eqnarray*}
Hence,  $I\left(\pi^{(a_{i+1},\ldots,a_L,a_1,\ldots,a_{i})};q^{(Lk+i)}\right)$
decreases strictly to zero as $k \to\infty$.
\end{proof}

\begin{corollary}[PCGS]
Suppose every $f_{a_i|b_i}$ is derived from the same joint pdf $f$, hence, $\{f_{a_i|b_i}:1\leq i\leq L\}$ is compatible, then as $k\to\infty$,
\[q^{(Lk+j)}\to \left\{
\begin{array}{ll}
f, & \mbox{ if $f_{a_{j}|b_{j}}$ is a full conditional (i.e., $c_j = D$),}\\
f_{c_{j}}, & \mbox{ otherwise.}
\end{array}\right.\]
\end{corollary}

\subsection{Mutually stationary and orthogonal distributions}
Though they reside on different $\CC_{c_i}$,
the stationary distributions of Algorithm~\ref{alg:2} are related. \citet{Kuo2019} called them \emph{pseudo-Gibbs distributions}, but never formally defined their mutual relationship.
\begin{definition}
For  $\{f_{a_i|b_i}: 1 \le i \le L\}$, assume that
$\order{1,\ldots,L}$ is a permissible updating cycle.
A collection of distributions,
$\{\pi^{(a_{i+1},\ldots,a_L,a_1,\ldots,a_i)} \in {\CC}_{c_i}: 1 \le i \le L\}$, are said to be mutually stationary when
$\II_{i}^{i+1}(\pi^{(a_{i+1},\ldots,a_L,a_1,\ldots,a_i)})
= \pi^{(a_{i+2},\ldots,a_L,a_1,\ldots,a_{i+1})}$ for every $i$,
with $(L+1)\equiv 1$.
\end{definition}
The stationary distributions produced by Algorithm~\ref{alg:2} are mutually stationary and they satisfy the following properties:
\begin{enumerate}[(i)]
\item Each set of  mutually stationary distributions are
associated with a specific permissible updating cycles, say $\order{1,\ldots,L}$.
\item Every $\pi^{(a_{i+1},\ldots,a_L,a_1,\ldots,a_i)}$ is stationary (the fixed point) with
 respect to the composite operator
 \[\II_{i-1}^{i}  \cdots \II_{i+1}^{i+2} \II_{i}^{i+1}.\]
\item Different permissible updating cycles have different sets of fixed points.  For example, the two sets of fixed points for $d=3$ and $\{f_{i|-i}\}$ are $\{\pi^{(1,2,3)}, \pi^{(2,3,1)}, \pi^{(3,1,2)}\}$ for $\order{1,2,3}$  and $\{\pi^{(1,3,2)},\\ \pi^{(3,2,1)}, \pi^{(2,1,3)}\}$ for $\order{1,3,2}$.
\item By the nature of conditional replacement operator, neighboring stationary distributions share the same $b_{i+1}$-marginal pdf:
\[\pi^{(a_{i+1},\ldots,a_L,a_1,\ldots,a_i)}_{b_{i+1}} = \pi^{(a_{i+2},\ldots,a_L,a_1,\ldots,a_{i+1})}_{b_{i+1}}.\]  When matching marginal holds for every $i$, Algorithm 2 has reached convergence, hence, it can be used as a convergence criteria.
\item For a compatible PCGS derived from a joint pdf $f$, the stationary distribution in every $\CC_{c_i}$ always exists.   Because  $\II_{i}^{i+1} (f_{c_i})= f_{c_{i+1}}$ for every $i$, distributions   $\{f_{c_i}, 1 \le i \le L \}$ are  mutually stationary.
\item When the stationary pdf has been reached in one $\CC_{c_i}$, one round of additional conditional replacements will bring about every other stationary distributions for this updating cycle.  This property has implication on the rate of convergence.
    For example, when $\pi^{(1,2,3)}$ is known, $\pi^{(2,3,1)}= \II_3^1(\pi^{(1,2,3)})$.
\item When  $c_i = D$ for every $i$, the mutually stationary distributions, $\{\pi^{(a_{i+1},\ldots,a_L,a_1,\ldots,a_i)}\}$, will all be joint pdfs.  If these joint pdfs are the same, then the conditional model is deemed compatible.
\item (Discrete case!!) Regardless of compatibility, if $S= S_1 \times \cdots\times S_d$, the mutually stationary distributions of $\{f_{i|-i}: 1 \le i \le d \}$ are known to exist, see \citet{Chen2015} and \citet{Kuo2019}.
\end{enumerate}
Therefore, we first determine a permissible updating cycle,
 then Algorithm 2 will approximate the $L$ mutually stationary distributions provided they exist.

\section{Examples}
\begin{example}\label{ex:1}
Consider $\{f_{1|23},f_{2|13},f_{3|12}\}$ where
\[f_{1|23}\sim N\left(\frac{-3X_2-X_3}2,1\right),\ f_{2|13}\sim N\left(\frac {-X_1-X_3}2,1\right),\]
\[\mbox{and }f_{3|12}\sim N\left(\frac {-3X_1-3X_2}2,1\right).\]
\citet{Li2012} used this model to show that  permissible updating cycle $X_1\to X_2\to X_3$ does not have a stationary distribution, but $X_2\to X_1\to X_3$ does. 
Using GS with ``$\cdots\to X_2^{(n)}\to X_1^{(n)}\to X_3^{(n)}\to X_2^{(n+1)}\to X_1^{(n+1)}\to\cdots $,'' we harvest the following three limiting distributions in batches of
$(X_1^{(n)},X_2^{(n)},X_3^{(n)})$, $(X_1^{(n)},X_2^{(n+1)},X_3^{(n)})$ and $(X_1^{(n+1)},X_2^{(n+1)},X_3^{(n)})$:

\[\pi^{(2,1,3)}\sim N_3\left(\left[\begin{array}{c}0\\ 0\\ 0\end{array}\right],\left[\begin{array}{ccc}241/50&-103/50&-207/50\\ -103/50&89/50&21/50\\ -207/50&21/50&329/50\end{array}\right]\right),\]
\[\pi^{(1,3,2)}\sim N_3\left(\left[\begin{array}{c}0\\ 0\\ 0\end{array}\right],\left[\begin{array}{ccc}241/50&-17/50&-207/50\\ -17/50&89/50&-61/50\\ -207/50&-61/50&329/50\end{array}\right]\right),\]
\[\pi^{(3,2,1)}\sim N_3\left(\left[\begin{array}{c}0\\ 0\\ 0\end{array}\right],\left[\begin{array}{ccc}241/50&-103/50&-73/50\\ -103/50&89/50&-61/50\\ -73/50&-61/50&329/50\end{array}\right]\right).\]
\begin{enumerate}[(i)]
\item The matching two-dimensional marginal pdf among neighboring stationary joint pdfs informs us that the convergence has been reached. In addition, $\II_3^2( \pi^{(2,1,3)})=\pi^{(1,3,2)}$ and $\II_2^1( \pi^{(1,3,2)})=\pi^{(3,2,1)}$
 imply that they are mutually stationary distributions.
\item Because the three stationary distributions are different, the conditional model is incompatible.
\item Though the  updating cycle $\order{1,2,3}$ is permissible, the GS does not converge to any  stationary distribution.
\item For an incompatible conditional model, especially in continuous cases, not every permissible updating cycle has a stationary distribution.  Thus, the assumption of existence of stationary distributions in Theorem~\ref{thm:convergence} is required.
\end{enumerate}
\end{example}

\begin{example}\label{ex:2}
Consider the conditional model for $(x_1,x_2,x_3)$: $\{f_{1|2},f_{2|3},f_{3|1}\}$ without a full conditional pdf, where
\[f_{1|2}\sim N\left(X_2/5,18/5\right),\ f_{2|3}\sim N\left(-5X_3/16,135/16\right),\]
and
\[f_{3|1}\sim N\left(-3X_1/4,55/4\right).\]
There is only one permissible updating cycle: $\order{1,3,2}$.   The three mutually stationary distributions  are
\[\pi_{23}^{(1,3,2)}\sim N_2\left(\left[\begin{array}{c}0\\ 0\end{array}\right],\left[\begin{array}{cc}10 & -5\\ -5&16\end{array}\right]\right),\]
\[\pi_{12}^{(3,2,1)}\sim N_2\left(\left[\begin{array}{c}0\\ 0\end{array}\right],\left[\begin{array}{cc}4&2\\ 2&10\end{array}\right]\right),\]
and
\[\pi_{13}^{(2,1,3)}\sim N_2\left(\left[\begin{array}{c}0\\ 0\end{array}\right],\left[\begin{array}{cc}4&-3\\ -3&16\end{array}\right]\right).\]
Since there is no stationary joint pdf, compatibility is automatic.   Moreover, if the joint pdf is assumed to be a trivariate Gaussian, then the joint pdf is uniquely determine by the three two-way marginal pdfs, see \citet{Wang1997}, and it is shown below:
\[N_3\left(\left[\begin{array}{c}0\\ 0\\ 0\end{array}\right],\left[\begin{array}{ccc}4& 2& -3\\ 2& 10&-5\\-3 &-5 &16\end{array}\right]\right).\]
\end{example}

\section{Concluding remarks}
In the past, people use one Markov operator, $K=T_1 \cdots T_d$, or one conditional expectation operator, $E=\EE[\EE[ \cdots \EE [\bullet \mid\FF_1]\cdots ]]$, to justify the GS.
Lumping multiple operators into one does simplify the theoretical framework, including turning a heterogenous chain into a homogenous chain.
But the simplifications offer little computational insight and fail to handle PCGS.
Instead,  we examine the effect of each $f_{a_i|b_i}$  separately, and invent the conditional replacement operator, which offers the following advantages:
\begin{enumerate}[(i)]
\item Iterative conditional replacements (ICR) require no background knowledge about Markov chain or conditional expectation, hence, makes the convergence easier to understand, especially for data scientists.
\item Algorithm 2 proves the convergence in terms of the Kullback-Leibler divergence, which implies the convergence in terms of $\|\bullet\|_{tv}$.
\item For dependence network, there is no need to expand every non-full conditional pdf into a full conditional pdf.
\item The proof of the convergence of PCGS suggests that $\II_i^{j}$ can be a replacement of individual transition kernel when studying heterogenous Markov chains.
\end{enumerate}

\section*{Acknowledgments}
The work of Kun-Lin Kuo was supported in part by the National Science and Technology Council, Taiwan (NSTC 112-2118-M-390-002).

\end{document}